\documentclass[12pt]{iopart}
\usepackage{graphicx,color,ulem}
\usepackage{amssymb,bm,iopams}

\renewcommand{\Re}{\mathop{\mathrm{Re}}}
\renewcommand{\Im}{\mathop{\mathrm{Im}}}
\renewcommand{\Tr}{\mathop{\mathrm{Tr}}}

\begin{document}

\title{Atom-number squeezing and bipartite entanglement of two-component
Bose-Einstein condensates: analytical results}
\author{G. R. Jin $^*$, X. W. Wang, D. Li, and Y. W. Lu}
\address{Department of Physics, Beijing Jiaotong
University, Beijing 100044, China} \ead{grjin@bjtu.edu.cn}

\begin{abstract}
We investigate spin dynamics of a two-component Bose-Einstein
condensates with weak Josephson coupling. Analytical expressions of
atom-number squeezing and bipartite entanglement are presented for
atom-atom repulsive interactions. For attractive interactions, there
is no number squeezing; however, the squeezing parameter is still
useful to recognize the appearance of Schr\"{o}dinger's cat state.

\vskip 0.5cm

\end{abstract}

 \maketitle



Atom-number squeezing has attracted much attention due to its
potential applications in quantum metrology and quantum information
\cite{Orzel,Greiner,Strabley,Esteve,Chuu,Jo}. As a special case of
spin squeezed states
\cite{Kitagawa,Wineland1,Wineland2,Hald,Kuzmich,Sorensen,Wang,Geremia,Jin09},
the number squeezed state shows reduced spin fluctuation of the
$J_z$ component below the standard quantum limit (SQL), which in
turn suppresses the deleterious effects arising from phase diffusion
\cite{Wright,Imamoglu,Javanainen,Castin,Law98,Vardi}. Dynamical
generation of the number squeezing has been theoretically
investigated based upon a two-mode boson model with a weak Josephson
coupling \cite{Bigelow,Law,Dunningham,Jin07,Jin08,Grond}.

So far, the number squeezing has been demonstrated indirectly by
detecting an increased phase fluctuation \cite{Orzel,Jo}, but not
atom-number variance. However, the variances of relative number and
phase has a nontrivial relation so a direct measurement of the
number fluctuation is of interests and necessary \cite{Jin08}. In
this letter, we study atom-number squeezing and bipartite
entanglement of a two-component BEC with the Josephson coupling. Two
analytical expressions are discovered, which provides us a more
direct way to measure the number variance and entropy of
entanglement through extracting phase coherence (i.e., the
visibility) in atomic interference experiments.


We consider a two-component Bose-Einstein condensate with hyperfine
states $|1\rangle $ and $|2\rangle$ coupled by an external microwave
(or rf) field \cite{Hall,Stenger,Schumm,BVHall}. For the BEC
confined in a deep three-dimensional harmonic potential, we adopt
single-mode approximation \cite{Milburn,Smerzi,Cirac,Vardi01}, which
results in the following Hamiltonian ($\hbar =1$):
\begin{equation}
H=\delta J_{z}-\Omega J_{x}+\chi J_{z}^{2},  \label{H}
\end{equation}%
where the detuning $\delta$, the Josephson-like coupling constant
$\Omega$, and the mean-field interaction strength $\chi$ can be
controlled artificially. Angular momentum operators
$J_{+}=(J_{-})^{\dag }=a_{2}^{\dag }a_{1}$ and $J_{z}=(N_2-N_1)/2$
are introduced, where $a_{\sigma}$ and $N_{\sigma}$
($=a_{\sigma}^{\dag} a_{\sigma}$) are the annihilation and number
operators for the two modes $\sigma=1$, $2$. Eq.~(\ref{H}) is
equivalent with a two-site Bose-Hubbard hamiltonian \cite{Folling},
where $\delta$ denotes potential bias of a double well, $\Omega$ the
hopping term, and $\chi$ the on-site interaction. Total particle
number $N=N_1+N_2$ is assumed to be a fixed c number, so atomic
number operators $N_{\sigma}=j+(-1)^{\sigma}J_{z}$ with $\sigma=1,
2$ and $j=N/2$. Atom number fluctuations, defined as usual $(\Delta
N_{\sigma})^{2}=\langle N_{\sigma}^{2}\rangle -\langle
N_{\sigma}\rangle^{2}$ are the same and equal to the variance
$(\Delta J_{z})^{2}$.

For any state vector $|\Psi \rangle$, one can determine the mean
spin: $\langle \mathbf{J}\rangle =(\langle J_{x}\rangle ,\langle J
_{y}\rangle ,\langle J_{z}\rangle )$, where $\langle J _{x}\rangle
=\Re\langle J_{+}\rangle $ and $\langle J _{y}\rangle =\Im\langle
J_{+}\rangle$. The expectation value $\langle J_{+}\rangle$ relates
to the first-order cross correlation function \cite{Law98,Vardi}:
\begin{equation}
g^{(1)}=\frac{|\langle a_{2}^{\dag }a_{1}\rangle |}{\sqrt{%
\langle N_1\rangle \langle N_2\rangle }}=\frac{|\langle J_{+}\rangle
|}{\sqrt{j^{2}-\langle J_{z}\rangle ^{2}}}, \label{PC}
\end{equation}
which measures phase coherence of the two-component BEC. It is
observable by extracting the visibility of atomic interference
fringes
\cite{Orzel,Greiner,Strabley,Esteve,Chuu,Jo,Schumm,Folling,Artur}.
Similar definition of the phase coherence has been proposed in Refs.
\cite{Meystra,Lee}. The degree of atom-number squeezing is
quantified by a parameter \cite{Esteve,Bigelow}
\begin{equation}
\xi ^{2}=\frac{N(\Delta N_{\sigma})^{2}}{\langle N_1\rangle \langle
N_2\rangle }=\frac{2j(\Delta J_{z})^{2}}{j^{2}-\langle J_{z}\rangle
^{2}},\label{xi2}
\end{equation}%
where total particle number $N=2j$ is assumed to be fixed. It has
been shown that there is no squeezing (i.e., $\xi^2=1$) for coherent
spin state (CSS) \cite{CSS}:
\begin{equation}
|\theta ,\phi \rangle=\exp [i\theta (J_{x}\sin \phi -J_{y}\cos
\phi)]|j,j\rangle, \label{CSS}
\end{equation}
which yields $\langle J_z\rangle=j\cos\theta$ and $(\Delta
J_{z})^{2}=(j/2)\sin^2\theta$. The condition of the number squeezing
is therefore $\xi^{2}<1$ \cite{Bigelow}, which is consistent with
previous one \cite{Jo}: $(\Delta J_{z})^{2}<j/2$ for a
\emph{symmetric} BEC with population imbalance $\langle J_{z}\rangle
=0$.

Firstly, let us consider spin dynamic of the symmetric BEC
($\delta=0$) with repulsive interactions ($\chi>0$) for an initial
CSS state $|\Psi(0)\rangle=|\frac{\pi}{2}, 0\rangle=|j, j\rangle_x$,
which is also an eigenvector of $J_x$ with eigenvalue $j$.
Experimentally, such a state has been prepared by applying a
two-photon $\pi/2$ pulse to the condensed atoms occupied in the
internal state $|2\rangle$ \cite{Hall}. The state vector at any time
$t$ can be expanded as $|\Psi\rangle=\sum_{m}c_{m}\vert j,m\rangle$,
where the probability amplitudes $c_{m}$ are determined by
time-dependent Schr\"{o}dinger equation with the initial condition:
$c_{m}(0)=\langle j,m|j, j\rangle_{x}=\frac{1}{2^{j}}{2j \choose
j+m}^{1/2}$. Note that the initial state shows the population
imbalance $\langle J_z(0)\rangle=0$ and the number variance $(\Delta
J_{z}(0))^{2}=j/2$ (i.e., $\xi^2=1$).

As the simplest case, we consider the Hamiltonian (\ref{H}) with
$\Omega=0$. This is the one-axis twisting model, proposed originally
by Kitagawa and Ueda \cite{Kitagawa}. The phase coherence can be
solved exactly as $g^{(1)}(t)=\cos ^{2j-1}(\chi t)$. In the
short-time limit, it decays exponentially as $g^{(1)}(t)\sim\exp
[-(t/t_{d})^{2}]$ with $\chi t_{d}=j^{-1/2}$, denoting a
characteristic time scale for phase coherence
\cite{Wright,Imamoglu,Javanainen,Castin}. The damping of phase
coherence, known as phase diffusion
\cite{Wright,Imamoglu,Javanainen,Castin,Law98,Vardi} has been
observed in experiment by extracting the visibility of the Ramsey
fringe \cite{Artur}. During phase diffusion, the number squeezing
$\xi^{2}$ remains constant because of conserved probability
distribution $|c_{m}|^{2}$.

The interplay between the nonlinear interaction and the Josephson
coupling leads to suppressed phase diffusion due to the appearance
of number squeezing. So far, the number squeezing has been
demonstrated for the BEC in optical lattices
\cite{Orzel,Greiner,Strabley,Esteve}, optical trap \cite{Chuu}, and
atom chip \cite{Jo}. To understand how this works, we consider
unitary evolution of the symmetric BEC governed by Hamiltonian
(\ref{H}). In Schr\"{o}dinger picture, the amplitudes always satisfy
the relation $c_{-m}=c_{m}$, which in turn leads to $\langle
J_{y}\rangle =\langle J_{z}\rangle =0$ and $\langle J _{x}\rangle
=\langle J_{+}\rangle \neq 0$. Moreover, Eq. (\ref{PC}) and Eq.
(\ref{xi2}) now reduce to $g^{(1)}=|\langle J_{x}\rangle |/j$ and
$\xi^{2}=2\langle J_{z}^{2}\rangle/j$. A relation between $g^{(1)}$
and $\xi^{2}$ can be obtained by examining Heisenberg equations of
motion:
\begin{eqnarray}
\dot{J}_{x} &=&-\chi (J_{y}J_{z}+J_{z}J
_{y}),   \label{Heisenberg1}\\
\dot{J}_{y} &=&\Omega J_{z}+\chi (J_{x}J_{z}+
J_{z}J_{x}),  \label{Heisenberg2} \\
\dot{J}_{z} &=&-\Omega J_{y}. \label{Heisenberg3}
\end{eqnarray}%
For time-independent $\Omega$ and $\chi$, we have $dJ
_{z}^{2}/dt=(\Omega /\chi)dJ_{x}/dt$, and thus
\begin{equation}
\xi^{2}(t)=1-\frac{2\Omega }{\chi}\left[1\mp g^{(1)}(t)\right] ,
\label{exact}
\end{equation}%
with the upper sign for the mean spin $\langle J_x\rangle\geq 0$ and
the lower one for $\langle J_x\rangle<0$. Eq. (\ref{exact}) provides
us an \emph{exact} relation between the number squeezing $\xi^{2}$
and the phase coherence $g^{(1)}$. To confirm it, we consider
two-particle ($N=2$) case. It is easy to obtain
$g^{(1)}=1-\frac{1}{2}(\chi\sin\omega_{2}t/\omega_{2})^{2}$ and
$\xi^{2}=1-\Omega \chi(\sin\omega_{2}t/\omega_{2})^{2}$, with
$\omega_{2}=(\Omega ^{2}+\chi ^{2}/4)^{1/2}$. Obviously,
Eq.(\ref{exact}) holds for the two-particle case. In addition, we
find that local minimum of $\xi^{2}$ (i.e., maximal number
squeezing) occurs at time $t_{\min }=\pi /(2\omega_{2})$. If an {\it
optimal} coupling $\Omega/\chi=1/2$ is applied, the system will
evolve into maximally number-squeezed state (MSS):
$|\Psi\rangle_{\mathrm{ MSS}}=ie^{-i\pi/(2\sqrt{2})}|1,0\rangle$,
which exhibits perfect squeezing $(\Delta J_{z})^{2}=0$
\cite{Wineland1}.

\begin{figure}[t]
\begin{center}
\scalebox{0.8}{\includegraphics[angle=0]{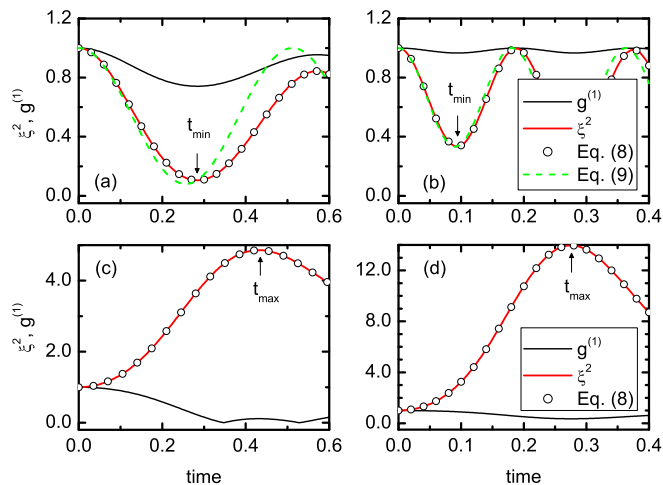}}
\end{center}
\caption{(Color online) Time evolution of phase coherence $g^{(1)}$
(black thin) and number-squeezing parameter $\xi^{2}$ (red) for
$N=20$, and the coupling: $\Omega/|\chi|=1.732$ (a) and (c),
$\Omega/|\chi|=10$ (b) and (d). In (c) and (d), negative $\chi$ case
is considered. The circles and the dashed green line are given by
Eq.~(\ref{exact}) and Eq.~(\ref{xi2ana}), respectively. The time is
in units of $|\chi|^{-1}$.} \label{fig1}
\end{figure}

For large $N$ case, there exist no exact solutions; however, some
approximated solutions are obtainable if the coupling is strong
enough. In this case, the mean spin $\langle J_{x}\rangle$ almost
remains unchanged at $j$. As a result, we adopt frozen-spin
approximation (FSA) \cite{Wineland1,Law,Aranya}, i.e., replacing
$J_{x}$ by $j$ ($=N/2$) in the Heisenberg equations, and obtain $
\ddot{J}_{z}=-\Omega ^{2}J_{z}-\Omega \chi
(J_{x}J_{z}+J_{z}J_{x})\simeq -\omega _{N}^{2}J_{z}$, where
$\omega_{N}=\sqrt{\Omega(\Omega +N\chi)}$ \cite{Law}. Solving the
above equation, it is easy to get harmonic solutions of $J_y(t)$ and
$J_z(t)$ \cite{Aranya}. Inserting them into Eq. (\ref{Heisenberg1}),
we obtain $J_{x}(t)$ and also $g^{(1)}=|\langle J_{x}\rangle
|/j\simeq 1-\frac{N}{2}(\chi \sin \omega _{N}t/\omega_{N})^{2}$.
Now, the phase coherence $g^{(1)}(t)$ almost exhibits sinusoidal
oscillations [thin lines of Fig.~\ref{fig1}], but not exponential
damping as previous case. In other words, the phase diffusion is
strongly suppressed due to the Josephson coupling. From
Eq.~(\ref{exact}), we further obtain
\begin{equation}
\xi^{2}(t)\simeq 1-N\Omega \chi \left(\frac{\sin
\omega_{N}t}{\omega_{N}}\right) ^{2}.  \label{xi2ana}
\end{equation}
Clearly, the maximal squeezing appears at time $t_{\min }=\pi
/(2\omega_{N})$ with, $\xi_{\min}^{2}=\Omega ^{2}/\omega_{N}^{2}<1$
\cite{Law,Aranya}. Dunningham et al. \cite{Dunningham} has
independently derived the time $t_{\min}$ using a semiclassical
analysis of Hamiltonian (\ref{H}) \cite{Smerzi}. For large $N$ case
($>10^3$), the optimal coupling obeys the power rule
$\Omega/\chi\sim 0.58 N^{0.32}$ \cite{Jin07,Jin08}. In
Fig.~\ref{fig1}(a), time evolution of $g^{(1)}$ and $\xi^2$ is
plotted for the optimal coupling $\Omega/\chi=1.732$ and $N=20$. The
FSA works (green dashed curves) quite well to predict $t_{\min}$.
For relatively large $\Omega/\chi$, say $\Omega/\chi=10$ for $N=20$,
the FSA follows full evolution of $\xi^2$ [see Fig.~\ref{fig1}(b)].
As shown Fig.~\ref{fig1}(c) and (d), Eq.(\ref{exact}) keeps hold for
negative $\chi$ case [see below].

\begin{figure}[th]
\begin{center}
\scalebox{0.85}{\includegraphics[angle=0]{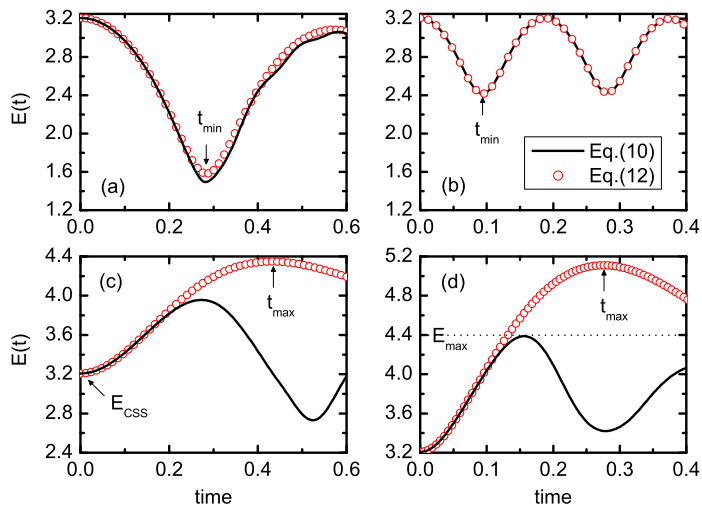}}
\end{center}
\caption{(color online) The entropy of entanglement $E$ (black solid
lines) for $N=20$. The red circles are given by Eq.~(\ref{Etxi2})
with numerical result of $\xi^2$. The initial value of $E$,
$E_{\mathrm{CSS}}=\frac{1}{2}\log[e\pi N/2]=3.21$, and the maximal
value $E_{\max}=\log(N+1)=4.39$. Other parameters are the same with
Fig.~\ref{fig1}.} \label{fig2}
\end{figure}

\begin{figure}[hbtp]
\begin{center}
\scalebox{0.5}{\includegraphics[angle=0]{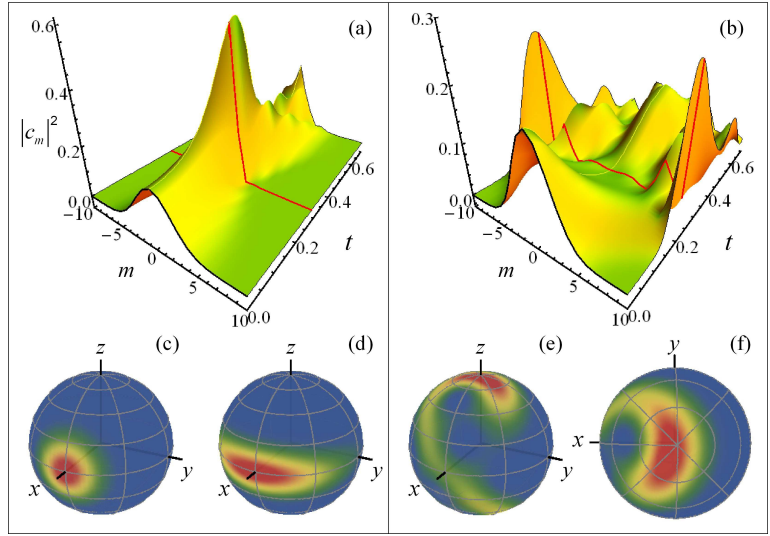}}
\end{center}
\caption{(color online) Probability distribution $|c_m|^2$ (a)-(b),
and quasi-probability distribution $Q(\theta, \phi)=|\langle \theta,
\phi |\Psi(t)\rangle|^{2}$ (c)-(f), where the CSS $|\theta,
\phi\rangle$ is given in Eq.~(\ref{CSS}). The red lines in (a) and
(b) represents $|c_m|^2$ for the MSS at $t_{\min}=0.284\chi^{-1}$
and the cat state at $t_{\max}=0.275|\chi|^{-1}$, respectively. The
parameters in the left and the right panels are the same with
Fig.~\ref{fig1}(a) and Fig.~\ref{fig1}(d).} \label{fig3}
\end{figure}

The relation between spin squeezing and quantum entanglement is of
interests. It was shown that the obtained squeezing is useful for
quantum metrology \cite{Wineland1,Wineland2} and entanglement
\cite{Sorensen}, provided that $\zeta _{\mathrm{S}}=\sqrt{2j}(\Delta
\hat{J}_{z})/|\langle \hat{J}_{+}\rangle|=\xi/g^{(1)}<1$
\cite{Esteve}. For the system considered here, however, only
entanglement between the two modes (i.e., bipartite entanglement) is
accessible due to the indistinguishability of identical bosons
\cite{Hines}. A standard measure of bipartite entanglement is the
so-called entropy of entanglement \cite{Hines,Nielsen}:
\begin{equation}
E(t)=-\Tr[\rho _{1}\log(\rho
_{1})]=-\sum_{m=-j}^{j}|c_{m}|^{2}\log\left(|c_{m}|^{2}\right),
\label{Entanglement}
\end{equation}%
where $\rho_{1}=\Tr_{2}(\rho)$ is the reduced density operator for
mode $1$ obtained by partial trace over mode $2$. The value of $E$
varies between $0$, for the separable product states to a maximum of
$\log(d)$, for maximally entangled states
$|\Psi\rangle_{\mathrm{MES}}=d^{-1/2}\sum_{m}|j,m\rangle$, where
$d=(2j+1)$ is the dimension of the Hilbert space. Utilizing
$c_{m}=d^{-1/2}$, Eq.~(\ref{Entanglement}) gives $E_{\max}=\log
(d)$, which is the maximum value of the entanglement for the system
\cite{Hines,Nielsen}. As an ansanz, the spin state $|\Psi
(t)\rangle$ can be treated as a Guassian \cite{Imamoglu}:
\begin{equation}
|c_{m}|^{2}\simeq \frac{1}{[2\pi (\Delta J_{z})^{2}]^{1/2}}\exp
\left[ -\frac{m^{2}}{2(\Delta J_{z})^{2}}\right] ,  \label{cm2}
\end{equation}%
with its width $(\Delta J_{z})=(j/2)^{1/2}\xi$, determined by the
number squeezing parameter $\xi^2(t)$. Substituting Eq.~(\ref{cm2})
into Eq.~(\ref{Entanglement}), and replacing the discrete sum over
$m$ by an integral, we arrive at
\begin{equation}
E(t)\simeq\frac{1}{2}\log\left[e\pi j \xi^2(t)\right], \label{Etxi2}
\end{equation}%
which provides us analytical relation between the number squeezing
and the two-mode entanglement. Considering $\xi^{2}=1$ for the
initial CSS $|j, j\rangle_x$, we have $E(0)=E_{\mathrm{CSS}}\simeq
\frac{1}{2}\log[e\pi j]$. Numerically, Hines et al. \cite{Hines}
have found that $E_{\mathrm{CSS}}/E_{\max}$ remains finite for large
$N$. Our result shows $E_{\mathrm{CSS}}/E_{\max}\rightarrow
\frac{1}{2}$ as $N\rightarrow \infty$. In Fig.~\ref{fig2}(a) and
(b), we plot time evolution of the entropy for finite $N=20$ case.
One can find that Eq.~(\ref{Etxi2}) (red circles) agrees very well
with exact numerical solution of Eq.~(\ref{Entanglement}). In the
initial stage, $E(t)$ decreases from $E_{\mathrm{CSS}}$ to its local
minimum at $t_{\min}$, due to the appearance of the MSS. In
comparison with the CSS, the MSS approaches to localized Twin-Fock
state \cite{TF}: $\vert N/2\rangle_{1}\vert N/2\rangle_{2}=|j,
0\rangle$, which exhibits $g^{(1)}=\xi^2=E=0$. To confirm it, we
plot evolution of probability distribution $|c_m|^2$ in
Fig.~\ref{fig3}(a). For the optimal coupling $\Omega/\chi=1.732$ and
$N=20$, the system evolves into the MSS after a duration $\chi
t_{\min}=0.284$, which shows probability distribution peaked at
$m=0$. Quasi-probability distribution $Q(\theta, \phi)$ of the
initial state is isotropic [Fig.~\ref{fig3}(c)], representing the
minimal uncertainty relationship of a coherent state
\cite{Kitagawa}. It becomes an elliptic shape due to the squeezing
along $J_z$ and the anti-squeezing along $J_y$ [Fig.~\ref{fig3}
(d)].

Until now, we consider spin dynamics of Hamiltonian ~(\ref{H}) with
$\chi>0$ and the initial state $|j, j\rangle_{x}$. This state is a
stable fixed point in phase space \cite{Boukobza}, so the phase
coherence $g^{(1)}$, the number squeezing $\xi^2$, and the two-mode
entanglement $E$ oscillate regularly with the same period. Using
Eq.~(\ref{exact}) and  Eq.~(\ref{Etxi2}), it is possible to measure
both $\xi^2$ and $E$ by extracting $g^{(1)}$ (i.e., the visibility)
in atomic interference experiment
\cite{Orzel,Greiner,Strabley,Esteve,Chuu,Jo,Schumm,Artur}. There are
two alternative experimental setups. One possibility is to trap the
condensed $^{23}$Na atoms in a symmetric double-well potential
formed by atom chip \cite{Jo}, the other is the two-component BEC in
an optical dipole trap \cite{Stenger}. In both cases, positive and
large enough $\chi\sim(a_{11}+a_{22}-2a_{12})$ is required, which
speeds up dynamics of the system such that the deleterious effects
like atom losses can be neglected \cite{Dunningham}.

To proceed, let us consider another scenario: spin dynamics of
Hamiltonian (\ref{H}) with negative $\chi$ case \cite{phase
sepration}. Now, the initial state $|j, j\rangle_x$ corresponds to
an unstable point at the separatrix \cite{Boukobza}, which leads to
a more complex dynamics with a quite different characteristic time
scale. From Eq.~(\ref{exact}), we find that there is no number
squeezing for negative $\chi$ case [see also Fig.~\ref{fig1}(c) and
(d)]. Instead of preparing the MSS, the latter model can be used to
generate Schr\"{o}dinger's cat state likes, $|\Psi\rangle_{\mathrm
{CAT}}= \frac{1}{\sqrt{2}}(|j, -j\rangle+ |j, j\rangle)$
\cite{mpe2}. In Fig.~\ref{fig2}(c) and (d), we plot time evolution
of $E(t)$ for $N=20$ case. Our results show that Eq.~(\ref{Etxi2})
follows exact results of $E$ in the early stages of the evolution,
then diverges as
$|\Psi\rangle\rightarrow|\Psi\rangle_{\mathrm{MES}}$. This is
because population distribution of $|\Psi\rangle_{\mathrm{MES}}$ is
no longer to be a Gaussian, and Eq.~(\ref{Etxi2}) can not simulate
the entropy. However, right-side of Eq.~(\ref{Etxi2}), a monotonic
function of $\xi^2$, is still useful to recognize the appearances of
the cat state, which exhibits the largest number variance $(\Delta
J_{z})_{\max}^{2}=j^2$ and $\xi_{\max}^2=2j$. In real evolution, an
approximate cat state with $\xi^2\sim1.3j$ [Fig.~\ref{fig1}(d)] can
be obtained at a time $|\chi| t_{\max}\sim\ln(8N)/N$ for the optimal
coupling $\Omega/|\chi|=N/2$ \cite{mpe2}. From right panel of
Fig.~\ref{fig3}, we also find that this state shows probability
distribution $|c_m|^2$ peaked at $m=\pm j$ and maximal values of
$Q(\theta, \phi)$ pointed to the north and the south poles of the
Bloch sphere.

In summary, we have investigated spin dynamics of a symmetric BEC
with repulsive interactions ($\chi>0$) evolved from a coherent spin
state $|j, j\rangle_x$. As main results of our work, we find
analytical expressions of the number-squeezing parameter $\xi^{2}$
and the entropy of entanglement $E$ as, Eq.~(\ref{exact}) and
Eq.~(\ref{Etxi2}). Both of them can be, in principle, at least,
measured by extracting the phase coherence $g^{(1)}(t)$ (i.e., the
visibility) in atomic interference experiments. For the case of
attractive interactions ($\chi<0$), thought there exists no number
squeezing, the squeezing parameter $\xi^{2}$ or Eq.~(\ref{Etxi2}) is
still useful to recognize the appearance of Schr\"{o}dinger's cat
state.

\ack

This work is supported by the NSFC (Contract No.~10804007), the
SRFDP (Contract No.~200800041003), and Research Funds of Beijing
Jiaotong University (Grants No.~2007XM049).




\begin{thebibliography}{99}
\bibitem{phase sepration} Phase seperation of the two-component BEC occurs for negative $\chi$ case, which can be avoided
by considering positive $\chi$ and the intial state $|j,
-j\rangle_x$ (cf. Ref. \cite{Hines}). Our results still hold.
\end{thebibliography}
\end{document}